\documentclass[proof]{WileyASNA-v1}

\articletype{Article Type}%
\usepackage{amsmath}

\received{26 April 2016}
\revised{6 June 2016}
\accepted{6 June 2016}

\begin{document}

\title{Towards the global magnetic field of the planet-hosting red giant eps Tau}

\author[1]{S.I. Plachinda*}

\author[1]{V.V. Butkovskaya}

\author[1]{N.F. Pankov}

\authormark{PLACHINDA \textsc{et al}}

\address[1]{\orgdiv{Stellar Physics Department}, \orgname{Crimean Astrophysical Observatory of RAS}, \orgaddress{\state{Nauchny}, \country{Russia}}}

\corres{*S. I. Plachinda, Crimean Astrophysical Observatory of RAS, Nauchny 298409, Russia. \email{psi1951@yahoo.com}}


\abstract{We present the results of a search for the magnetic field inhomogeneity for the red giant $\epsilon$ Tau. This research is based on observations obtained over 10 nights in 2008-2010 with the ESPaDOnS CFHT spectropolarimeter. We found a previously undescribed instrumental effect in the ESPaDOnS spectra, consisting in random polarization outliers. Therefore, to measure the magnetic field from the unblended individual lines, we preliminarily cleared the initial array of spectral lines from the lines distorted by polarization outliers. On only one date from ten, the magnetic field of $\epsilon$ Tau was found to exceed 3$\sigma$. We also revealed that during two nights the time series of the magnetic field values shows a distribution that is different from the normal distribution. A hypothesis was put forward that this may be due to the inhomogeneity of the magnetic field of this star.}


\jnlcitation{\cname{%
\author{S.I. Plachinda}, 
\author{V.V. Butkovskaya}, 
\author{N.F. Pankov}} (\cyear{2020}), 
\ctitle{Towards the global magnetic field of the planet-hosting red giant eps Tau}, \cjournal{AN}, \cvol{2016;00:1--6}.}

\fundingInfo{}

\maketitle


\section{Introduction}\label{sec1}

$\epsilon$ Tau (Sp G9.5 III, HR 1409, HD 28305) is a weakly active red giant belonging to the open cluster of Hyades. Its mass is $M = 2.5 - 2.7~M_\astrosun$, radius is $R = 13.7~R_\astrosun$, X-ray luminosity is $L_\mathrm{x} = 2\times10^{28}~\mathrm{erg~s^{-1}}$, and $\epsilon$ Tau has a massive planet ($M_\mathrm{p} = 7.1 - 7.6~ M_\mathrm{J}$) orbiting it with a period of 594.9 $\pm$ 5.3 days  ~\citep{ Arentoft2019, Auriere2015, Gondoin1999, Sato2007}.

\cite{Sato2007} obtained an effective temperature $T_\mathrm{eff} = 4901 \pm 20$ K, a surface gravity log $g = 2.64 \pm 0.07~\mathrm{cm~s^{-2}}$, a microturbulent velocity $v_\mathrm{t} = 1.49 \pm 0.09~\mathrm{km~s^{-1}}$, and a metallicity $[\mathrm{Fe/H}] = 0.17 \pm 0.04$, which is consistent with the mean metallicity of the Hyades. ~\cite{Gray1982} derived a projected rotational velocity of the star $v \sin i = 2.5~\mathrm{km~s^{-1}}$.

\cite{Auriere2015} detected a weak magnetic field of $B_\mathrm{e} \sim -1.3 \div 1.4$ G for  $\epsilon$ Tau via Zeeman signatures revealed by them during the spectropolarimetric monitoring of a number of giant stars with the twin spectropolarimeters ESPaDOnS at the Canada-France-Hawaii Telescope (CFHT) and Narval at Télescope Bernard Lyot (TBL, Pic du Midi Observatory) over 11 nights in 2008 – 2010.

To measure the magnetic field, the authors used the LSD (Least Square Deconvolution) technique, which calculates the mean Stokes $V$ and $I$ profiles over all spectral lines, including blends ~\citep{Donati1997}.

The LSD method usually uses the most complete array of spectral lines (including blends) to reconstruct pseudo-mean Stokes profiles, which makes it possible to dramatically increase the signal-to-noise ratio in the case of a sun-like spectrum and to register weak magnetic fields up to one tenth of Gauss. This method was used to carry out extensive observational campaigns to measure magnetic fields in stars of different spectral classes and luminosity types. In the last decade, LSD profiles have been actively used to construct magnetic maps of stellar surfaces using Zeeman-Doppler imaging (ZDI). Certainly, the original LSD method and its later modifications revolutionized the technique of studying the weak magnetic fields of stars.

If we want to achieve maximum signal-to-noise ratio, the LSD method is preferable (for example, to detect weak magnetic fields), but if our goal is to investigate inhomogeneous physical conditions on the surface of the star, the Single Line (SL) method ~\citep{Butkovskaya2007, Plachinda2004, Plachinda1999} is more suitable. Different spectral lines in the stellar atmosphere can be formed under different physical conditions, both on the surface and with depth. Therefore, measuring the magnetic field using the pseudo average spectral line profiles can lead to the distortion of information about the geometry of the magnetic field ~\citep{Plachinda2019}.

For high-precision measurements of the magnetic field on the Sun as a star, single spectral lines are usually used, and different lines give different values of the magnetic field ~\citep{Stenflo2013}. It is also known that different lines in the solar spectrum can have different asymmetries ~\citep{Sheminova2020, Dravins2008}. These effects can be caused both by different depth of spectral lines formation and by the formation of spectral lines in different areas of supergranules. These facts are true for all stars with convective envelopes. If there are also different types of active regions on the surface of the star, this, in addition, complicates the physical conditions in which the spectral lines are formed. Unlike the LSD method, the SL method requires a higher signal-to-noise ratio in the observed spectra, because it usually uses fewer suitable spectral lines.

Modern large telescopes allow us to measure the stellar magnetic field using single lines, thereby making it possible to study in detail the physics of the magnetic field, including revealing subtle patterns in the structure of the magnetic field.

\begin{figure*}[t]
	\centerline{\includegraphics[width=390pt,height=39pc]{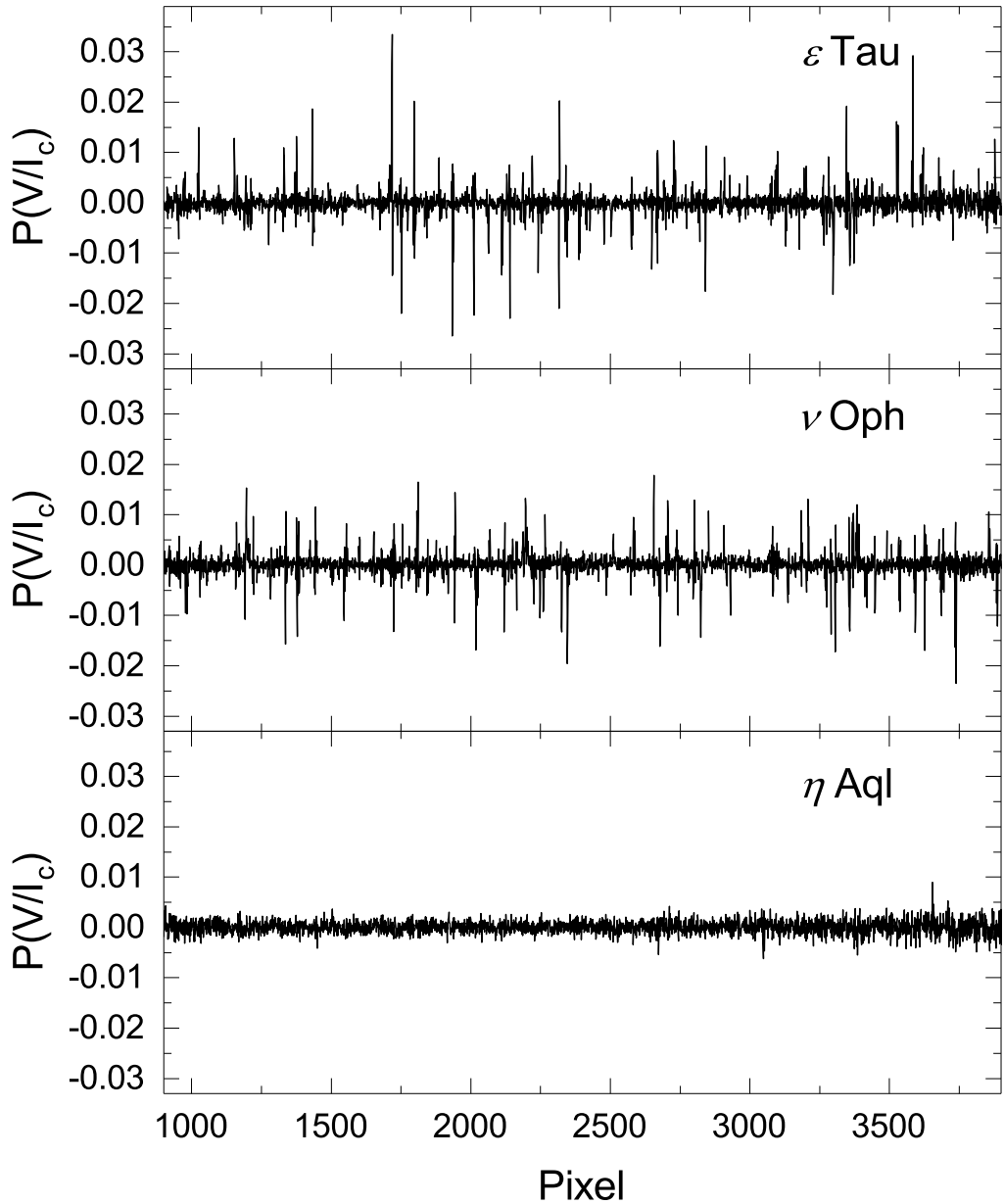}}
	\caption{Circular polarization in the spectrum of $\epsilon$ Tau (2008-10-18), $\nu$ Oph (2008-10-18), and $\eta$ Aql (2017-08-08). The Y-axis shows the Stokes V profile, and the X-axis shows the pixel numbers. The edges of the spectra on the blue and red sides are cut off due to the low signal-to-noise ratio.\label{fig1}}
\end{figure*}

Our SL-method (the center of gravity method for single lines) calculates the longitudinal magnetic field from individual spectral lines using four polarized spectra normalized to the continuum, obtained in two consecutive exposures at different (orthogonal) angular positions of the input quarter-wave plate:

	\begin{multline}
		\label{field}
	B_e = k((\lambda_{1r} - (\lambda_{2l} + \Delta\lambda_{ins}))/2 - \\
	(\lambda_{1l} - (\lambda_{2r} + \Delta\lambda_{ins}))/2)/2 = \\
	k((\lambda_{1r} - \lambda_{2l})/2 + (\lambda_{2r} - \lambda_{1l})/2 )/2 = k\Delta\lambda_B,
	\end{multline}
	
where $\lambda_{1r}$ and $\lambda_{1l}$ - the wavelength of the center of gravity of the line in  right and left polarized spectra in the first exposure; $\lambda_{2r}$ and $\lambda_{2l}$ - the wavelength of the center of gravity of the line in right and left polarized spectra in the second exposure; $\Delta\lambda_{ins}$ - the instrumental shift; $k = 1/(4.67\times 10^{-13}z\lambda^2)$, where $z$ is the Lande factor.

The normalized to continuum circular polarization (Stokes $V$) can be written as

	\begin{multline}
		\label{polariz}
	V/I_{c} = ((I_{1r}-(I_{2l} + \Delta I_{ins}))/2 - 
	(I_{1l}-(I_{2r} + \Delta I_{ins}))/2)/2 = \\
	((I_{1r}-I_{2l})/2 + (I_{2r}-I_{1l})/2)/2.
	\end{multline}
	
The center of gravity of each polarized component of the spectral line is calculated with using cubic spline interpolation or least-squares cubic spline fitting  ~\citep{Calif2015}. In the case of statistically homogeneous array, the entire array of magnetic field measurements, $N = N_{pair~ of~ exposures} \cdot N_{lines}$, is used to calculate mean per night magnetic field and its standard error. To test the statistical homogeneity of the resulting array of magnetic field measurements, the Monte Carlo method under the assumption of a normal probability distribution is used. If the time series of the measured magnetic field is uniform, the difference between the experimental standard error and the Monte Carlo simulated standard error is only a few percent: see section  \textit{4. The Reliability of the “Flip-Flop” Zeeman Measurement Technique} in ~\cite{Plachinda2004}. The array of magnetic field measurements can become statistically inhomogeneous for a number of reasons: unaccounted instrumental effects, variability of the magnetic field during the observation night, inhomogeneity of the magnetic field in the photosphere, inhomogeneity of physical conditions on the surface of the star, and so on.
	
Despite the fact that the measured magnetic field for most of the weakly active giants from the sample of~\cite{Auriere2015} is close to zero, we selected $\epsilon$ Tau, for which there are 10 nights of observations with the ESPaDOnS spectropolarimeter and on some nights a magnetic field was registered, to search for the signatures of inhomogeneity of its magnetic field using SL technique. Another red giant, $\nu$ Oph (Sp G9 III, HR 6698, HD 163917), was taken as a control object for which no magnetic field was detected. The stellar parameters of $\nu$ Oph, $T_\mathrm{eff} = 4831$ K, $M = 3.0~ \mathrm{M_\astrosun}$ ~\citep{Auriere2015, Sato2012}; $L_\mathrm{x} = 7\times10^{28}~\mathrm{erg~s^{-1}}$ ~\citep{Gondoin1999}, are close to the stellar parameters of $\epsilon$ Tau.

\begin{center}
	\begin{table*}[t]%
		\caption{Magnetic field of $\epsilon$ Tau and $\nu$ Oph \label{tab1}}
		\centering
		\begin{tabular*}{500pt}{@{\extracolsep\fill}lccccccccccD{.}{.}{11}c@{\extracolsep\fill}}
			\toprule
			\textbf{Date} & \textbf{HJD}  & \textbf{$B_\mathrm{e}^{*}$}  & \textbf{$\sigma_\mathrm{B}^{*}$}  & \textbf{$B_\mathrm{e}$}  & \textbf{$\sigma_\mathrm{B}$}  & \textbf{ND$_\mathrm{Be}$}  & \textbf{$B_\mathrm{null}$}  & \textbf{$\sigma_\mathrm{Bnull}$}  & \textbf{ND$_\mathrm{Bnull}$}  & \textbf{$N_\mathrm{t}$/$N_\mathrm{s}$} \\
			
			\text{UT} & \text{(2450000+)}  & \text{G}  & \text{G}  & \text{G}  & \text{G}  & \text{}  & \text{G}  & \text{G}  & \text{}  & \text{} \\
			
			\midrule
			$\epsilon$ Tau \\	
			
			2008-08-23 & 4702.02  &  1.40  & 0.58 &  0.80  & 1.43  & Y &  3.22  & 1.39 & Y & 784/599  \\
			2008-10-18 & 4757.94  &  0.68  & 0.45 &  5.48  & 1.56  & Y &  0.97  & 1.28 & Y & 785/508  \\
			2008-12-17 & 4817.80  & $-$1.34  & 0.27 &  0.04  & 0.85  & N &  1.02  & 0.78 & Y & 789/691  \\
			2009-10-02 & 5107.01  &  0.61  & 0.34 & $-$2.75  & 2.17  & Y & $-$1.01  & 1.58 & Y & 774/306  \\
			2009-10-07 & 5112.01  &  0.59  & 0.48 & $-$0.33  & 1.75  & Y &  1.69  & 1.56 & Y & 765/431  \\
			2010-03-08 & 5263.72  &  0.97  & 0.38 & $-$2.78  & 2.30  & Y & $-$3.52  & 1.72 & Y & 697/275  \\
			2010-07-20 & 5398.12  & $-$0.52  & 0.35 & $-$1.07  & 1.81  & Y &  1.76  & 1.48 & Y & 707/366  \\
			2010-10-19 & 5489.16  & $-$1.00  & 0.32 & $-$2.30  & 2.04  & Y & $-$1.01  & 1.83 & Y & 604/261  \\
			2010-11-16 & 5517.02  & $-$0.59  & 0.23 & $-$2.40  & 1.20  & N &  2.13  & 0.92 & Y & 755/466  \\
			2010-11-22 & 5522.86  &  0.02  & 0.34 & $-$3.27  & 1.89  & Y &  1.15  & 1.60 & Y & 624/310  \\
			$\nu$ Oph  \\
			2008-08-24 & 4702.74  & $-$0.98  & 0.64 & $-$0.61  & 1.46  & Y & $-$2.88  & 1.39 & Y & 781/523  \\
			2008-10-18 & 4757.69  &  1.02  & 0.58 &  0.43  & 1.50  & Y & $-$1.75  & 1.33 & Y & 781/484  \\
			
			\bottomrule
		\end{tabular*}
	\end{table*}
\end{center}

\section{Data reduction and treatment}\label{sec2}

Spectropolarimetric observations of $\epsilon$ Tau and $\nu$ Oph were performed at CFHT ESPaDOnS over 10 nights in 2008 – 2010 and over 2 nights in 2008, respectively. From 2006 to 2011A, ESPaDOnS was equipped with an EEV CCD42-90-1-941 detector with 2k × 4.5k 0.0135 mm square pixels. Since 2011A, EEV CCD42-90-1-941 has been replaced by a new detector E2V CCD42-90-1-B32 named Olapa (see, for example, ~\cite{Wade2015}).

Each Stokes $V$ exposure sequence consists of four subexposures that are recorded at the following consecutive positions of the entrance quarter-wave plate: $+45^{\circ}$, $-45^{\circ}$, $-45^{\circ}$, $+45^{\circ}$. The integration and read-out time during one subexposure is 32 s and 25 s in 2008 and 60-70 s and 40 s in 2009-2010, respectively. The typical signal-to-noise ratio of a single spectrum is 390 in 2008 and 500-600 in 2009-2010. The resolving power of spectra is $R=68~000$. For the magnetic field calculation, we used spectral lines in the range 4100-7000 \AA.

The reduction and calibration of these spectra were performed using the standard IRAF software. In order to avoid the distortion of the circular polarization in line profiles during continuum normalization, the continuum functions of each order of the spectra were calculated using the Fortran code developed by S. Plachinda. 

This procedure according to formula (\ref{polariz}) allows us to control the presence of polarization artifacts in the spectra normalized to the continuum. The examples of such polarization artifacts in the spectrum after normalization to the continuum is shown in Fig. 1 for $\epsilon$ Tau (top panel) and $\nu$ Oph (middle panel). The absence of such polarization outliers for $\eta$ Aql is illustrated in bottom panel of Fig. 1. All three stars were observed with ESPaDOnS.

The upper and middle panels of Fig. 1 show circular polarization artifacts in the spectra of the same order for the stars $\epsilon$ Tau and $\nu$ Oph obtained on 2008-10-18. The amplitude of these polarization outliers reaches several percent. Observations were obtained with the EEV1 detector and similar polarization artifacts were detected by us in all orders on all observation nights.

A cross-correlation analysis of arrays of polarization values obtained for both stars from observations on 2008-10-18 was performed. The analysis showed the absence of any correlations. Thus, we supposed the process to be sporadical. Obviously, this effect cannot be caused by the optics of the polarimeter and spectrograph. Consequently, it could be caused by sporadical malfunctions in the electronics of the CCD camera. The bottom panel of Fig. 1 shows a similar diagram of circular polarization for the classical Cepheid $\eta$ Aql for comparison. The star was observed at ESPaDOnS in 2017 with the new E2V CCD detector (Olapa). There are no circular polarization artifacts in this graph.

It should be noted that we did not detect polarization artifacts from observations in 2006 (EEV1 detector) of the slowly rotating magnetic star 33 Lib ~\citep{Butkovskaya2019}. We also found no such polarization artifacts in observations of the classical magnetic star $\beta$ CrB with other telescopes: over 32 nights in 1993-2004 with the coude long-slit spectrograph at the 2.6 m Shajn reflector of the Crimean Astrophysical Observatory, and over 6 nights in 2007-2009 with the high- resolution echelle spectropolarimeter BOES at the Bohyunsan Optical Astronomy Observatory ~\citep{Han2018}.

\begin{figure*}[t]
	\centerline{\includegraphics[width=342pt,height=22pc]{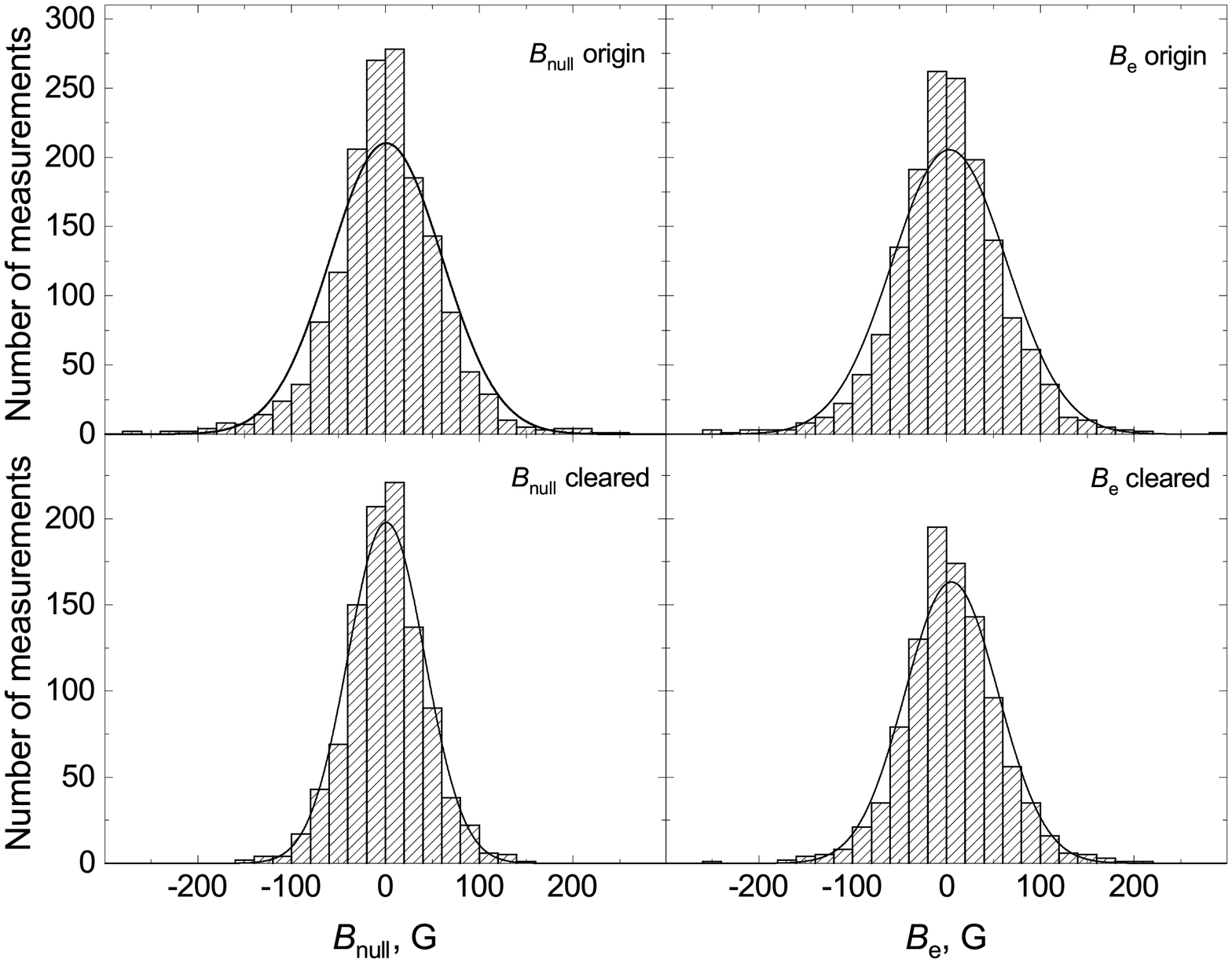}}
	\caption{Distribution of the \textquotedblleft{null}\textquotedblright~field and the magnetic field of $\epsilon$ Tau measured from individual spectral lines on the date 2008-10-18. The normal distribution curves are shown by solid lines. In the upper panels, \textquotedblleft{$B_\mathrm{{null}}~\mathrm{origin}$}\textquotedblright~is a distribution of the \textquotedblleft{null}\textquotedblright~field  measured over all the unblended spectral lines, \textquotedblleft{$B_\mathrm{{e}}~\mathrm{origin}$}\textquotedblright~is a distribution of the magnetic field of $\epsilon$ Tau measured over all the unblended spectral lines. In the bottom panels, \textquotedblleft{$B_\mathrm{{null}}~\mathrm{cleared}$}\textquotedblright~is a distribution of the \textquotedblleft{null}\textquotedblright~field measured with the array of spectral lines, cleared from the lines distorted by outliers, \textquotedblleft{$B_\mathrm{{e}}~\mathrm{cleared}$}\textquotedblright~is a distribution of the magnetic field measured with the array of spectral lines, cleared from the lines distorted by outliers.\label{fig2}}
\end{figure*}

The calculation of the magnetic field longitudinal component (see formula (\ref{field}) above) for $\epsilon$ Tau and $\nu$ Oph has been performed by measuring the Zeeman splitting in individual spectral lines, using the procedure discussed in detail by ~\citep{Butkovskaya2007, Plachinda1999, Plachinda2004}. The SL-method (Single Line) developed at CrAO makes it possible to measure the magnetic field by the centers of gravity of the polarized components of individual spectral lines. For each date, its own \textquotedblleft{mask}\textquotedblright~is built, which includes a set of spectral lines and their parameters. It should be noted that the number of spectral lines in the \textquotedblleft{mask}\textquotedblright~can vary from date to date (see the last column in Table 1). This is due to the variation of the number of weak spectral lines, which are appropriate for inclusion their in the \textquotedblleft{mask}\textquotedblright. Then the resulting array of measurements of the magnetic field along individual lines is tested for uniformity. If the array is statistically  heterogeneous, then at the next step the possibility of dividing it into statistically significantly different subarrays of lines formed under similar physical conditions is investigated. The detection of two or more subarrays of lines giving reliably different values of the magnetic field may indicate an inhomogeneity of the magnetic field on the surface of the star ~\citep{Plachinda2019}.

The SL method uses some approaches to estimate the reliability of the obtained results (see Section 4 in ~\cite{Plachinda2004}:
\begin{enumerate}
	\item The control of polarization signal distortion and the detection of apparent outliers when the spectra are normalized to the continuum.
	\item The analyzing of the statistical distribution of magnetic field values.
	\item The checking of the coincidence between the Monte-Carlo simulated and experimental standard error using the normal distribution of a probability function.\\
	It is allowed to use the normal distribution, because when the read-out noise is neglected with  respect to a high signal-to-noise, the signal in each pixel of the CCD-detector is typically normal-distributed. Such an approach enables us to calculate the Monte-Carlo standard deviations for each single magnetic field measurement.
	\item Application of the homogeneous array of measurements for estimation of the mean and its standard error.
	\item Calculation of the “null” field is an internal spectropolarimeter test for presence of significant stochastic outliers or spurious wavelength- or time-dependent Stokes signatures.
	In the absence of instrumental effects, the “null” field should be statistically insignificant.
\end{enumerate}

\section{Results}\label{sec3}

Result of magnetic field measurements for $\epsilon$ Tau and $\nu$ Oph is shown in Table 1. The first and second columns contain the dates and Heliocentric Julian Dates of observations. The third and fourth columns show the mean magnetic fields and their errors from ~\cite{Auriere2015}.  The fifth and sixth columns present the magnetic fields and their errors obtained in this study. Column 7 indicates whether the distribution of the magnetic field (ND$_\mathrm{Be}$ - Normal Distribution) measured from the undistorted individual spectral lines is normal (Y – distribution is normal, N - distribution is not normal). The next three columns are the same as columns 5-7, but for the “null” field. The last column shows the initial number of unblended spectral lines ($N_\mathrm{t}$ - total) and the final number of spectral lines undistorted by polarization outliers ($N_\mathrm{s}$ - selected).

A preselection of the unblended lines in the spectral range 4100-7000 \AA~was performed using the appropriate temperature and gravity for each star from the data provided by the Vienna Atomic Line Database VALD ~\citep{Kupka1999}. Since a number of lines in the spectra of $\epsilon$ Tau and $\nu$ Oph were distorted by instrumental outliers, we excluded them from the total array of unblended spectral lines used to measure the magnetic field. To determine the presence of instrumental artifacts in the lines, we calculated a \textquotedblleft{null}\textquotedblright~field for each line. Ideally, the \textquotedblleft{null}\textquotedblright~field should not exceed $\pm3\sigma$, where the error of a single measurement, $\sigma$, is calculated by the Monte-Carlo method under the assumption of a normal distribution. In this case, we used the $\pm4\sigma$ criterion due to the possible presence of unaccounted effects that can lead to an increase in scatter of the field values. Having extracted an array of the \textquotedblleft{null}\textquotedblright~field for each spectral line, we excluded from further calculations those spectral lines for which this \textquotedblleft{null}\textquotedblright~field exceeded $\pm4\sigma$.

Analysis of the obtained arrays of the \textquotedblleft{null}\textquotedblright~field showed that for both stars on each date up to 50\% of the spectral lines give a field, the absolute value of which exceeds $4\sigma$ and in some cases reaches two dozen $\sigma$. Moreover, on different dates, a statistically significant \textquotedblleft{null}\textquotedblright~field was recorded on different spectral lines. We found no mention of this effect in the literature describing observations with  ESPaDOnS. It seems impossible to detect it using methods that involve automatic processing of observations and/or measurements of the magnetic field from averaged profiles.

Figure 2 shows histograms of the \textquotedblleft{null}\textquotedblright~field and the magnetic field distribution for $\epsilon$ Tau (2008-10-18) before cleaning the array from the distorted lines (top panel) and after cleaning (bottom panel). The histograms of the \textquotedblleft{null}\textquotedblright~field and magnetic field distributions before removing the  distorted spectral lines demonstrate a significant deviation from the normal distribution, while the arrays cleaned from the distorted lines exhibit the normal distribution with a confidence level higher  than 99.9\%. The arrays were tested for normal distribution using the Kolmogorov-Smirnov test ~\citep{JASP2020} for a significance level of 0.001 (0.1\%).

It should be noted that on 2 dates there is a deviation of  histograms of the magnetic field of $\epsilon$ Tau from the normal distribution (see Table 1). On these nights, 2008-12-17 and 2010-11-16, three and two sets of observations were carried out respectively, while on other dates, only one set of observations was conducted. One of the reasons for the deviation from the normal distribution may be the inhomogeneity of the magnetic field, which is difficult to detect based on a small number of measurements. However, further observations are required to confirm or refute this hypothesis.

We roughly estimated the discrepancy in accuracy between LSD and SL methods for $\nu$ Oph, i.e. determined the effect of random instrumental outliers on measurement accuracy. We did it as follows:

The measurement accuracy limit in this case is set by the SL Monte Carlo method assuming a normal distribution of the measured values of the magnetic field. According to the Monte Carlo method, the $\nu$ Oph magnetic field array is homogeneous. Therefore, we can use the standard errors from Table 1. \cite{Auriere2015} used from 6500 to 14000 spectral lines for LSD magnetic field calculation. We took minimal number of spectral lines $N_{LSD} = 6500$. The standard errors for the first and second dates are $\sigma_{1LSD} = 0.64$ G and $\sigma_{2LSD} = 0.58$ G. The number of spectral lines, which were used for SL-method, $N_{1SL} = 523$ and $N_{2SL} = 484$, as well as standard errors, are presented in Table 1. Using $N_{LSD} \approx N_{SL}\cdot(\sigma_{SL}/\sigma_{LSD})^2$ ~the number of spectral lines are roughly estimated as $N_{1(LSD)}=2722$ and $N_{2(LSD)}=3237$. Therefore, in this particular case, when the instrumental random outliers of polarization are absent, the LSD method needs in about 3000 lines instead of 6500 to achieve standard errors $\sigma_{1(LSD)} = 0.64$ G and $\sigma_{2(LSD)} = 0.58$ G.

\section{Conclusions}\label{sec5}
A search for the inhomogeneity of the magnetic field on the red giant $\epsilon$ Tau was carried out using spectropolarimetric observations acquired over 10 nights in 2008-2010 with the ESPaDOnS CFHT spectropolarimeter. The red giant $\nu$ Oph with the zero magnetic field, whose physical parameters are close to the physical parameters of $\epsilon$ Tau, was taken as a control star. We found some of the lines in spectra of both stars to be distorted by random polarization outliers, presumably of an instrumental nature. These lines were excluded from the general array of spectral lines used for calculating the magnetic field of stars. For $\epsilon$ Tau, a magnetic field exceeding 3$\sigma$ was recorded on one from ten nights. No statistically significant magnetic field on $\nu$ Oph was found. We also noticed that on two nights the statistical distribution of the magnetic field values of $\epsilon$ Tau shows a deviation from the normal distribution. We assume that this may be due to the inhomogeneity of the magnetic field. However, further observations are required to confirm or refute this hypothesis.


\section*{Acknowledgments}
We are grateful to the anonymous referee for a careful review of the manuscript, resulting in insightful recommendations and kind suggestions leading to its improvement. Plachinda S. acknowledge the support of \fundingAgency{Ministry of Science and Higher Education of the Russian Federation} under Grant number \fundingNumber{075-15-2020-780 (N13.1902.21.0039)}. This study is based on observations obtained at the Canada–France–Hawaii Telescope (CFHT). This research used the facilities of the Canadian Astronomy Data Centre operated by the National Research Council of Canada with the support of the Canadian Space Agency.

\nocite{*}
\bibliography{Plachinda}%

\section*{Author Biography}

\begin{biography}{}{\textbf{Sergei Plachinda} PhD, Leading Scientific Researcher at Crimean Astrophysical Observatory.}
\end{biography}

\end{document}